\documentclass[fleqn,usenatbib]{mnras}

\usepackage{newtxtext,newtxmath}
\usepackage{placeins}

\usepackage[T1]{fontenc}
\usepackage{ae,aecompl}

\usepackage{graphicx}   
\usepackage{amsmath}    
\usepackage{amssymb}    

\newcommand\msun{\rm M_{\odot}}
\newcommand\rsun{\rm R_{\odot}}
\newcommand\lsun{\rm L_{\odot}}
\newcommand\mdot  {\dot{M}}

\newcommand\kms{\rm \, km ~ s^{-1}}
\newcommand\msunyr{\rm M_{\odot}\,yr^{-1}}

\newcommand\beq{\begin{equation}}
\newcommand\eeq{\end{equation}}
\newcommand{\cmthree}{{\rm cm}^{-3}}

\title[FU Ori winds] {Disc wind models for FU Ori objects}
\author[Milliner et al.]
      {Kelly Milliner$^{1}$, James H. Matthews$^{2}$, Knox S. Long$^{3}$, Lee Hartmann$^{1}$
       \thanks{E-mail:
          lhartm@umich.edu}
\\
$^{1}$Department of Astronomy, University of Michigan,  500
           Church Street, Ann Arbor, MI 48105, USA \\
$^{2}$University of Oxford, Astrophysics, Keble Road, Oxford, OX1 3RH, UK\\
$^3$Space Telescope Science Institute, 3700 San Martin Drive, Baltimore, MD 21218, USA\\
}

\date{Accepted XXX. Received YYY; in original form ZZZ}

\pubyear{2018}

\begin{document}
\label{firstpage}
\pagerange{\pageref{firstpage}--\pageref{lastpage}} \maketitle

\begin{abstract}
We present disc wind models aimed at reproducing the main features of the strong Na I resonance line P-Cygni profiles in the rapidly-accreting pre-main sequence FU Ori objects. We conducted Monte Carlo radiative transfer simulations for a standard magnetocentrifugally driven wind (MHD) model and our own ``Genwind'' models, which allows for a more flexible wind parameterisation.
We find that the fiducial MHD wind and similar Genwind models,
which have flows emerging outward from the inner disc edge, and thus have polar cavities with no absorbing gas, cannot reproduce the deep, wide Na I absorption lines in FU Ori objects viewed at low inclination.  We find that it is necessary to include an ``inner wind'' to fill this polar cavity to reproduce observations.  In addition, our models
assuming pure scattering source functions in the Sobolev approximation at intermediate viewing angles ($30^{\circ} \lesssim i \lesssim 60^{\circ}$) do not yield sufficiently deep line profiles. Assuming complete absorption yields better agreement with observations, but simple estimates strongly suggest that pure scattering should be a much better approximation. 
The discrepancy may indicate that the Sobolev approximation is not applicable, possibly due to turbulence or non-monotonic velocity fields; there is some observational evidence for the latter.  Our results provide guidance for future attempts to constrain FU Ori wind properties using full MHD wind simulations, by pointing to the importance of the boundary conditions necessary to give rise to an inner wind, and by suggesting that the winds must be turbulent to produce sufficiently deep line profiles.

\end{abstract}




\section{Introduction} 
\label{sec:intro}

Jets and winds from young stellar objects are ubiquitous \citep[see the review by][]{frank14}. For low-mass, pre-main sequence stars, the the high-velocity winds and jets are clearly driven by disc accretion, as shown by the observational correlation between mass loss signatures and the infrared excesses produced by dusty circumstellar discs \citep{cabrit90,hartigan95} and the correlation between mass loss and accretion rates \citep{calvet98}. The low wind temperatures and weak ultraviolet radiation fields of these systems show that driving by thermal or radiation pressure is inadequate to account for the high-velocity winds and jets;  instead, magnetic acceleration is favored \citep{pudritz83} \citep[see][for a recent review]{turner14}, especially as acceleration by field lines anchored in the rotating disc naturally produces the strong collimation seen in jets \citep[e.g.,][]{konigl89,shu95}.

Exactly where the high-velocity outflows are launched in the low-mass, pre-main sequence T Tauri stars is complicated by the presence of an inner magnetosphere which truncates the disc, leading to competing models in which the jet/outflow either originates at the disc-magnetosphere interface \citep{shu94} or over a finite region of the (inner) disc \citep{pudritz83,konigl89,anderson03}. In addition, lower-velocity outflows are observed which are thought to emerge from still larger disc radii \citep[$\gtrsim 0.5$~AU;][]{hartigan95,msimon16}. Whether these slow winds are magnetically-driven is less clear, as photoevaporation mass loss may dominate \citep{clarke01,alexander04,gorti09,owen10,owen11}. 
The situation in the outer disc is complicated 
because non-ideal MHD effects come into play due to low levels of ionization, which can result in a wide range of complex behavior \citep[e.g.,][]{simon13,bethune17} depending upon several poorly-constrained parameters, including
dust size and spatial distributions, disc chemistry, and the amount of vertical magnetic flux penetrating the disc \citep{bai13,bai16,simon13}.


Given these complexities, it is worth investigating disc winds without the complications of magnetospheric truncation and non-ideal MHD.
The rapidly-accreting, pre-main sequence FU Ori objects \citep{herbig77,hartmann96} arguably provide examples of such winds.
At their high accretion rates of $10^{-5} - 10^{-4} \msunyr$, FU Ori disc temperatures and thus ionization states should be high enough that ideal MHD applies, with the
 magnetorotational instability (MRI) operating to drive accretion \citep{balbus98}.  Unlike the situation for T Tauri stars, there is no evidence for magnetospheric accretion in FU Ori objects \citep{hartmann11,hartmann16},
 consistent with theoretical expectations that high disc ram pressures should compress any magnetosphere back to the stellar photosphere.

\begin{figure}
     \centering
     \includegraphics[width=0.55\textwidth]{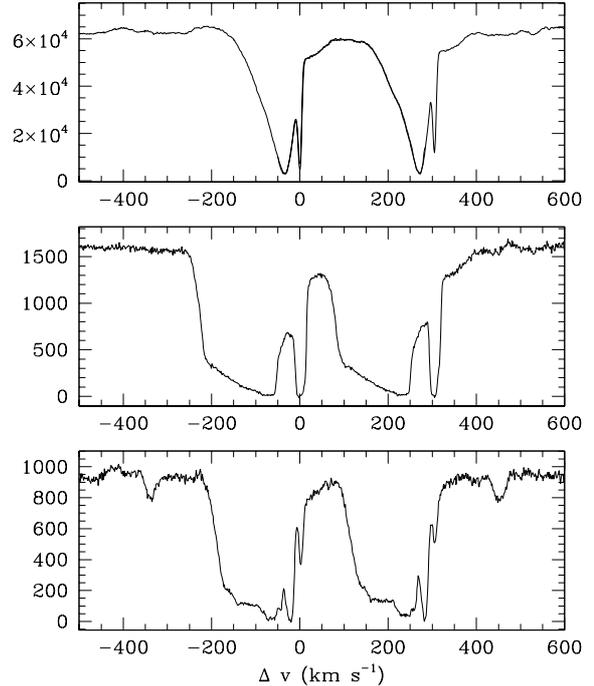}
     \caption{Na I spectra of (top to bottom) FU Ori, taken on 12/10/2011 (MJD 55905.48, V10157 Cyg, taken on 8/13/2008 (MJD 54691.49, and V1515 Cyg, also taken on 8/13/2008 (MJD 54691.47), all using the HIRES spectrometer on the Keck I Telescope, with exposure times of 15 minutes for FU Ori and 5 minutes each for the other two objects. The spectra have been shifted to place the narrow interstellar absorption feature at zero velocity (D1), as this represents the systemic velocities of these objects.}
     \label{fig:fuori}
 \end{figure}

 
As might be expected for objects with high accretion rates, FU Ori objects generally show evidence for strong winds in P Cygni profiles, especially in the Na I resonance lines
\citep{bastian85,croswell87,reipurth90} (see Figure \ref{fig:fuori}).
The extreme depth of Na absorption  demonstrates that these winds must cover the line of sight to nearly all of the optically-emitting disc surface, demonstrating that these flows are true disc winds \citep{calvet93}.
Moreover, observations of slightly blue-shifted intermediate-strength photospheric lines in FU Ori show that the wind is initially rotating with the
disc \citep{calvet93,hartmann95}.

Thus, the FU Ori objects provide unique opportunities to explore the properties of magnetic winds from turbulently-accreting discs.
However, the only theoretical modeling of strong lines that trace the high-velocity regions of FU Ori winds -- namely, H$\alpha$ and the Na I resonance doublet -- is that of \cite{croswell87}, who constructed detailed non-LTE models but with very simplistic wind geometries using either spherical or plane-parallel geometries.  Neither of these assumptions is useful for exploring geometrical constraints such as collimation of the outflow, dependence of mass flux on radius, etc.  

In this paper we
use a Monte Carlo radiative transfer code developed to model the winds from other accretion disc systems \citep{long02,higginbottom13,matthews15} to explore the implications of the observed Na I resonance line profiles of FU Ori objects for constraining wind properties.  We focus on the Na I lines in this preliminary exploration because they are strong enough to trace the flows to high velocities.
As the heating and thus the temperature structure of FU Ori winds is poorly understood, we do not solve for ionization balance; instead, in this preliminary investigation we assume a high enough abundance of Na I to make the lines very optically thick, consistent with observations, and concentrate on the resulting geometric constraints.

 We find that it is very difficult to explain the observed profiles with source functions assuming pure scattering, even though rough estimates suggest that this should be a good approximation. We suggest that the problem may be with the inapplicability of the Sobolev limit due to turbulence in the winds.  We also find that the structure of the inner wind (small cylindrical radii) is also important for explaining the line profiles, but exactly how to treat this region is uncertain because of our poor understanding of the inner boundary conditions of FU Ori discs.
Our results provide a starting point for further investigations using global MHD disc and wind simulations.

\section{Na I P Cygni profiles}
\label{sec:obs}

Figure \ref{fig:fuori} shows
examples of recently-observed Na I P Cygni line profiles for the three best-studied FU Ori objects.\footnote{The profiles for FU Ori and V1515 Cyg are fairly typical \citep{bastian85,croswell87,hartmann95}; V10157 Cyg has been more variable, as discussed below.} The first notable property of these profiles is the lack of significant redshifted absorption or emission, unlike what is seen for example in the P Cygni resonance line profiles of hot stars.
There is some weak absorption on the
long-wavelength side of the D$_2$ lines in FU Ori and V1057 Cyg (discussed further in \S 6), but these features do not extend to high velocities,
in contrast with the blueshifted absorption.
While this is consistent with ejection from an optically-thick disc which occults the expected red-shifted outflow, if the wind has a significant divergence, redshifted material can absorb light from the far side of the disc in inclined systems.\footnote{Essentially all the optical light comes from the disc and not the central star. There is no evidence for any stellar photospheric absorption lines, as expected since in outbursts the disc becomes $\sim 10^2$ times brighter than in the quiescent state, consistent with typical T Tauri parameters; \cite{herbig77}.}

The only significant constraint on wind opening angles among these three objects is provided by FU Ori
itself, as near-infrared interferometry suggests that the disc has a significant inclination of $i = 55^{\circ}$ to the line of sight \citep{malbet05,quanz06}.  Presuming that the wind is collimated along the disc rotation axis, the limited redshifted absorption in these systems suggests that the half-angle divergence of the flow must  $\lesssim 40^{\circ}$.  V1057 Cyg and 
V1515 Cyg do not provide strong constraints as they are probably observed much more pole-on.  If these objects have the same intrinsic rotational velocities as FU Ori, then the differences in the projected rotational velocities 
\citep{hartmann85} would suggest inclinations of about $24^{\circ}$ and $12^{\circ}$
for V1057 Cyg and V1515 Cyg, respectively, qualitatively
consistent with the 
the geometries of the wind-driven envelope cavities of these objects seen in scattered light \citep{goodrich87}, assuming the outflows are collimated perpendicular to the disc.

The second important observational constraint is the extreme depth of the blueshifted absorption component, reaching nearly zero residual intensity.
As pointed out by \cite{calvet93}, this means that, at the relevant blueshifted velocities, there must be optically-thick line absorption which intercepts the line of sight to essentially all of the radiating disc photosphere, which basically requires a wind emerging from a large region of the disc.

While we only consider steady wind models with smooth flows in this paper, it must be recognized that the observed line profiles are variable, with indications of distinct velocity components (e.g., the low-velocity absorption dip and other structure in the blueshifted absorption of V1515 Cyg in Figure \ref{fig:fuori}).  A particularly notable example of clumpy outflow has been seen in V1057 Cyg. In 1981, the continuous broad Na I absorption seen in Figure \ref{fig:fuori} was not present; instead, the blueshifted absorption was confined to a few weaker, narrow components running from about $-180$ to $-90 \kms$ \citep{bastian85}.  By 1985, the absorption collapsed to
a single narrow component at about $-90 \kms$ \citep{croswell87}.
This suggests that in the 1981-85 time frame the continuous wind was not detectable in Na I (though contemporaneous H$\alpha$ profiles showed a continuing wind), and the observed Na I absorption components were basically ``shells'' at large distances from the disc.
In general it seems likely that FU Ori winds are generally ``clumpy'',
with the impression of smooth flow enhanced when the Na I optical depths in the wind are large.

\begin{figure}
    \centering
    \includegraphics[width=0.35\textwidth]{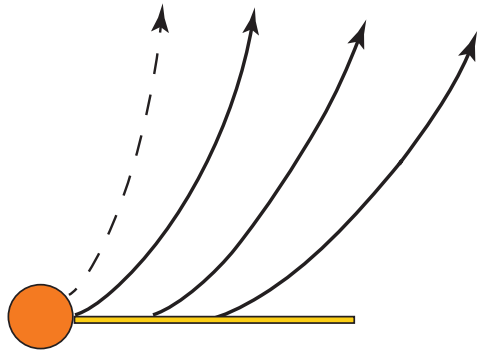}
    \caption{Schematic diagram for the  wind geometry. The solid curves indicate the flow for the standard wind models, for which there is no wind interior to the field line which connects to the disc at the stellar radius. The dashed curve refers to models which include a ``stellar'' wind.}
    \label{fig:diskwind_starx}
\end{figure}

\section{Wind models}

\label{sec:models}

\subsection{Magnetocentrifugally-driven (MHD) wind models}
\label{sec:mhdmodels}

The first type of wind models that we investigate are ``MHD'' models, which are based on the \cite{blandford82} (BP82) axisymmetric self-similar solutions for a magnetocentrifugally-driven, cold wind emanating from a thin disc. In these models, the wind is entirely driven by centrifugal acceleration of the magnetic field lines whose footpoints are fixed in the Keplerian disc.  For mass loss to occur it is necessary for the field lines to make an angle $\geq 30^{\circ}$ with respect to the rotation axis as they emerge from the disc; at larger radii the field lines generally converge toward the axis.  The basic setup is indicated in Figure \ref{fig:diskwind_starx}.

As in BP82, we use cylindrical coordinates 
\begin{equation}
[R, \phi, z] = [ R_f \xi, \phi, R_f \chi ] \,,
\end{equation}
such that magnetic field lines (streamlines) are self-similar as defined by non-dimensional functions $[\xi, \phi, \chi]$, scaled to the footpoint radius in the thin disc at $R = R_f ,z = 0$.
For simplicity and to control the amount of collimation directly, we do not solve the cross-field balance but simply impose streamlines of the form
\begin{equation}
\chi = a \xi^2 + b \xi - (a + b)\,,
\label{eq:fieldlines}
\end{equation}
where the final constant yields
$\chi = 0$ when $\xi = 1$ \citep[see further discussion in][]{lima10}.  The adopted quadratic form mimics the general nature of solutions for disc winds driven by rotating magnetic field lines, with an initial tilt away from the rotation axis for outward acceleration, with collimation toward the axis at larger distances due to hoop stresses.

Following BP82, the 
wind velocity components are given by
\begin{equation}
[v_R, v_{\phi}, v_z] = [\xi' f(\chi),g(\chi), f(\chi)] (G M/R_f)^{1/2}\,,
\end{equation}
where $f$ and $g$ are the similar velocities scaled to the Keplerian velocity at the field line footpoint at $R_f$.  Given
$a$ and $b$, and additional parameters $\kappa$, which is related to the ratio of the constant mass flux to the constant magnetic flux, and $\lambda$, which is a measure of the specific angular momentum of the flow (BP82), the structure of the flow along the streamline can be found by solving a quartic equation for the Alfv\'enic Mach number \citep[Equation 2.12 in BP82; see also][]{lima10}.

Finally, we derive mass volume densities $\rho$ from assuming steady flow, with a scaling at the disc $\rho \propto R^{-3/2}$.  Because $v_z (z = 0) = 0$,
we use the product
\begin{equation}
\rho v_z = (\rho v_z)(z=0) {(R_{in}/R_{f})^2 \over \xi (\xi - \chi \xi')}\,,
\label{eq:rhovz}
\end{equation}
where the product
\begin{equation}
(\rho v_z)(z=0) 
= { \mdot_w \over
4 \pi R_{in}^2 \ln (R_{out}/R_{in})}\,,
\label{eq:rhovzscale}
\end{equation}
Here $\mdot_w$ is the total disc mass loss rate and $R_{in}$ and $R_{out}$ are the inner and outer disc radii.  The density then follows once $v_z$ is determined.

\begin{figure}
 \centering
    \includegraphics[width=\columnwidth]{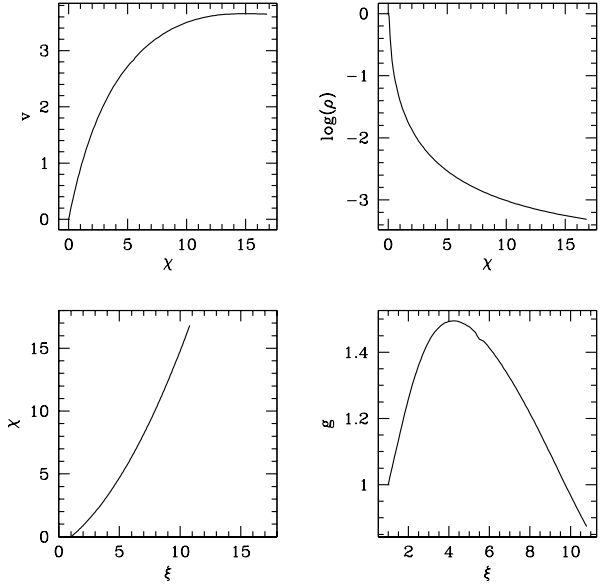}
    \caption{Example self-similar wind solution for $a = 0.0936, b=0.612, \kappa = 0.03$, and $\lambda = 30$. The streamline shape is shown in the lower left panel, while the poloidal velocity, density, and azimuthal velocity are shown in the upper left, upper right, and lower right, respectively. }
    \label{fig:bpstd}
\end{figure}

In Figure \ref{fig:bpstd} we show
a typical solution for the normalized velocities and densities adopting $a = 0.0936, b=0.612$, parameters chosen to approximately represent the streamlines shown in Figure 3 of BP82, along with $\kappa = 0.03, \lambda = 30$.
The magnetic field causes the flow to rotate at nearly constant angular velocity until just before the Alfv\'en point; the input of rotational energy drives the outflow to terminal velocities $\sim 3.2$ times the Keplerian
velocity at the field footpoint.
The Alfv\'en point is at $\xi \approx 5.5, \chi \approx 8$; at this position the normalized velocities are
$\xi' f = 1.11$, $g= 1.17$, $f= 2.73$, and
$\rho = 2.55e-3$.

\subsection{Genwind models}

\label{sec:genwindmodels}

The advantage of the MHD models is that the wind rotation is internally self-consistent with the acceleration of the outflow.  However, we found that it was not easy with our method to explore variations in outflow divergence.
Moreover, there are good reasons to question the applicability of BP82-type solutions in the case of FU Ori objects. The BP82 solutions assume that the disc is very thin, while accretion disc calculations indicate that
the inner discs are geometrically thick, with aspect ratios $H/R \sim 0.3$ \citep{clarke90,bell94}.  In addition,
the BP82 model assumes that the only accelerating force is that due to the rotating magnetic field lines; but because the disc is thought to be turbulent due to the MRI, one expects that magnetic waves propagating away from the disc could contribute strongly to driving the flow.  For these
reasons we developed another wind model with parameterized geometry and velocity laws (here called ``Genwind'' models for concision) to explore the effects differing amounts of collimation and acceleration on the line profiles.

We adopted
the same geometry for fieldlines (streamlines) as in Equation \ref{eq:fieldlines}, but replaced the
MHD solution with the following equation
for the vertical velocity,
\begin{equation}
v_z = \eta \left ({G M \over R_f} \right)^{1/2} \left ( 1 - {R_f^2 \over R^2 + z^2} \right )^{\zeta}\,,
\end{equation}
  The $R$-component of the velocity is then
$v_R = \xi' v_z$. In the absence of
an obvious alternative we simply assume constant angular momentum for the gas,
\begin{equation}
    v_{\phi} = \left ({G M \over R_f} \right)^{1/2} \left ( { R_f \over R}\right ) \,.
\end{equation}
The density is then determined from 
Equations \ref{eq:rhovz} and \ref{eq:rhovzscale} as in the previous case.

\begin{figure}
    \centering
    \includegraphics[width=0.4\textwidth]{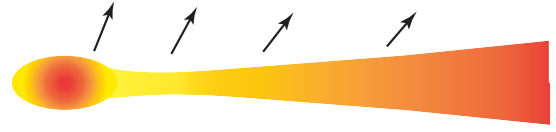}
    \caption{Cartoon of possible inner disc structure of FU Ori objects, showing the thick disc, which merges with the bloated star without forming a boundary layer.}
    \label{fig:cartoon_fuori}
\end{figure}

\subsection{Inner wind models}

\label{sec:innerwindmodels}
The models discussed in the previous
subsections assume that there is no wind interior
to the fieldline (streamline) emanating from the inner edge of the disc, at the stellar radius (Figure \ref{fig:diskwind_starx}).
As will be shown in \S \ref{sec:results},
winds with this geometry have difficulties in
reproducing the observations. We therefore constructed models with "inner winds" to fill in the polar holes in the outflow (schematically indicated by the dashed curve in Figure
\ref{fig:diskwind_starx}).

It is highly unlikely that any significant inner wind is a true or standard stellar wind, because the estimated mass loss rates imply an energy flux comparable to that produced by accretion \citep{calvet98}, and thus are roughly two orders of magnitude larger than the entire luminosity of the pre-accretion-outburst star.  Instead, any such outflow must be powered by accretion, and thus likely produced in some complex interaction region, of uncertain structure, where the inner disc merges with the star.
Models of the spectral energy distribution of FU Ori and other systems imply inner disc radii about twice that of typical T Tauri stars of similar mass \citep{kenyon88,zhu07}.
This suggests that the outer layers of the central stars have expanded due the advection of large amounts of thermal energy into their upper layers (\citealt{baraffe12}; see also \citealt{prialnik85}).  We conjecture that the inner regions look something like the cartoon in Figure \ref{fig:cartoon_fuori}, where the hot, geometrically-thick inner disc somehow merges onto
the bloated, rapidly-rotating outer stellar layers.

With this cartoon picture in mind we therefore treat the inner wind as an extrapolation of our Genwind models, such that
\begin{equation}
v_z = \eta \left ({G M \over R_{in}} \right)^{1/2} \left ( 1 - {R_f^2 \over R^2 + z^2} \right )^{\zeta}\,.
\label{eq:vzstellar}
\end{equation} 
The $R$-component of the velocity is again
$v_R = \xi' v_z$. 
The density is given by
\begin{equation}
\rho v_z = {\rho_0 v_z (0) \over \xi (\xi - \chi \xi')}\,.
\label{eq:rhovzstellar}
\end{equation}
These choices maintain continuity with the disc wind and avoid singularities at $R=0$.
Finally, we again assume constant
angular momentum for the gas assuming solid
body rotation of the object inside $R_0$ for lack of any better idea,
\begin{equation}
    v_{\phi} = \left ({G M \over R_f} \right)^{1/2} \left ( { R_f \over R}\right ) \,.
\end{equation}

The radiative transfer code does not implement a geometry 
like that shown in Figure \ref{fig:cartoon_fuori}; the disc is still a thin structure and the star is a sphere.  However, by setting the stellar effective temperature low enough, the star does not
contribute a significant amount of light at
6000 \AA\,. Thus the effect of the inner wind is to
extinct light from the disc.

\section{Calculations}

Line profiles were calculated for the Na I
resonance doublet using a modified version of the Monte Carlo radiative transfer code ``Python'' originally described by \cite{long02}.\footnote{More information about and the source code for Python can be found at https://github.com/agnwinds/python.} The code has been applied to disc winds in cataclysmic variables \citep{long02,noebauer10,matthews15}, active galactic nuclei \citep{higginbottom13,matthews16} and young stellar objects \citep{sim05} and is designed to simulate line and continuum transfer through an astrophysical plasma in the Sobolev approximation.  Since the Na I doublet lines are resonance lines, we treat them in a simple 2-level approximation. The line optical depths are calculated in the Sobolev approximation and this Sobolev optical depth is also sampled to calculate the scattering anisotropy. We conduct our simulations in 2.5D, in that we assume azimuthal symmetry and mirror symmetry through the disc plane, but track photons in 3D. The radiation sources in the system are the central star and the disc.  The temperature of the disc varies with radius according to the prescription for a standard steady-state disc.  Both are assumed to radiate as blackbodies. 

The heating and thus the temperature structure of FU Ori winds are poorly understood.  Therefore, although the code is capable of computing a fully consistent ionization balance, here we fix the ionization state of Na I with a high enough abundance to make the lines very optically thick.  This allows us to place limits on the wind geometry from the depths and shapes of the lines as a function of viewing angle and other wind parameters.  

The cylindrical grid is divided into $n_r \times n_z = 100 \times 100$ cells, 
with the coordinates $(r_i, z_j)$ sampled according to
\begin{align}
    r_i &= L_r 10^{(i-1) \log(R_{max}/(L_r n_r))},\\
    z_j &= L_z 10^{(j-1) \log(Z_{max}/(L_z n_z))},
\end{align}
where $i$ and $j$ are the indices of the cell, and $R_{max}$ and $Z_{max}$ are the extent in $r$ and $z$, respectively. The scale lengths, $L_r$ and $L_z$, were systematically varied between $1 \times 10^{11} $ cm and $5 \times 10^{7} $ cm until yielding converged profiles, which occurred at a value of $5 \times 10^{9} $ cm.

We adopt typical parameters for the disc as those from FU Ori itself, taken from the careful spectral energy modeling by \cite{zhu07}. The best-fitting models indicate $T_{max} = 6420$~K \citep[see also][]{zhu09}.
After accounting for the closer Gaia distance of 416 pc vs. the assumed 500 pc in \cite{zhu07}, we find
$L_{acc} = 157\, \lsun$, $R_{in} = 4.16 \, \rsun$,
and $M \mdot = 2.4 \times 10^{-5} \msun^2 yr^{-1}$.  Finally, using the observed rotational velocities
with an inclination of $i = 55^{\circ}$
\citep{malbet05,quanz06}, $M = 0.27 \msun$ and $\mdot = 1.5 \times 10^{-4} \msunyr$.  These parameters then yield the Keplerian velocity at the inner disc edge
of $112 \kms$.  We use the same disc
parameters to compare with the V1057 Cyg and V1515
Cyg observations; while these are undoubtedly somewhat different in reality, the spectral energy distributions can be fit with similar values of
$T_{max}$.  Moreover, while the true rotation rates are not well-determined because these objects are viewed roughly pole-on (\S \ref{sec:obs}), the rotation of the wind is not important for comparing the models with observations.

Because in outburst the optical luminosity of these discs is roughly 5 magnitudes higher than pre-outburst levels \citep[e.g.,][]{herbig77} as well as T Tauri stars, it is not surprising that there is no evidence of a stellar spectrum in the observed profiles.
Thus we initially set the central star effective temperature to a typical value $3800$~K, which satisfies observational constraints and means that the optical continuum emission at the wavelengths of the Na I resonance doublet arises entirely from the disc. 

Using a combination of constraints from prior modeling of the Na I resonance lines and H$\alpha$ \citep{croswell87} and intermediate-strength photospheric lines \citep{calvet93,hartmann95}, the mass loss rate of FU Ori has been estimated as $\sim 10^{-5} \msunyr$, consistent the ratio $\mdot_w/\mdot \sim 10^{-1}$ found for T Tauri stars \citep{calvet98}. We adopt that value in our models, but note that the resulting line profiles depend on the product $\mdot_{w} f_{abs}$, where $f_{abs}$ is the constant fractional abundance we assume for the population in the ground state of the Na I doublet, assuming a total solar Na abundance relative to hydrogen of $2 \times 10^{-6}$.  The models of \cite{croswell87} exhibited typical values
$f_{abs} \approx 10^{-4}$;
as the observed Na I absorption is very deep, we take a large value of $f_{abs} = 10^{-2}$ so as to concentrate on the geometrical constraints on the outflow needed to extinct as much of the disc radiation as possible (while avoiding producing unobserved damping wings).

We compute the profiles for source functions in the extreme limits of pure scattering and pure absorption.  While the former case is much more likely (see Discussion), we include pure absorption to make the absorption profiles as deep as possible for a given wind geometry.

\begin{table}
 \caption{Wind models}
 \label{tab:emodels}
 \begin{tabular}{lcccccc}
  \hline
Model & a & b & $\eta$ & $\zeta$ & Type & Inner wind? \\
  \hline
MHD1 & 0.936 & 0.612 & -- & -- & MHD & No \\
MHD2 & 0.936 & 0.612 & -- & -- & MHD, no $v_{\phi} $ & No \\
MHD1i & 0.936 & 0.612 & -- & -- & MHD & Yes \\
G3 & 0.08  & 0.4 & 1.5  & 2  & Genwind & No \\
G4 & 0.08  & 100 & 1.5  & 2  & Genwind & No \\
G5 & 0.08  & 0.4 & 1.5  & 2  & Genwind & Yes \\
G6 & 0.04  & 0.4 & 1.5  & 2  & Genwind & Yes \\
G7 & 0.01  & 0.4 & 1.5  & 2  & Genwind & Yes \\
  \hline
 \end{tabular}
\end{table}



\begin{figure*}
    \includegraphics[width=0.38\textwidth]{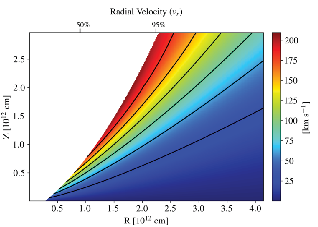}
    \includegraphics[width=0.38\textwidth]{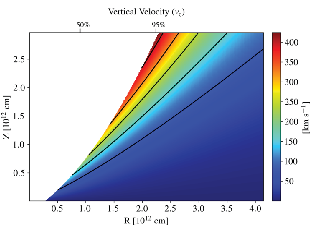}
    \caption{Radial (left) and vertical velocity field of the MHD Model 1.
    For reference we also mark the radial positions which encompass 50\% and 95\% of the total disc light at 5900 \AA.}
    \label{fig:mhdgrid1}
\end{figure*}

\begin{figure*}
    \includegraphics[width=0.38\textwidth]{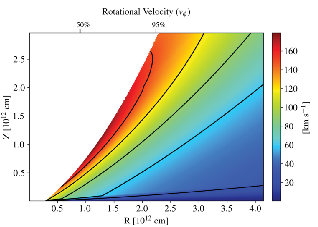}
    \includegraphics[width=0.38\textwidth]{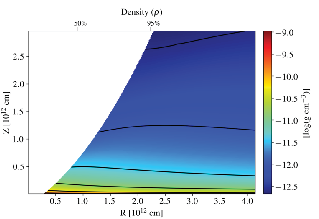}
    \caption{Rotational velocity and density fields of the MHD Model 1.}
    \label{fig:mhdgrid2}
\end{figure*}

\begin{figure*}
\centering
\includegraphics[width=\textwidth]{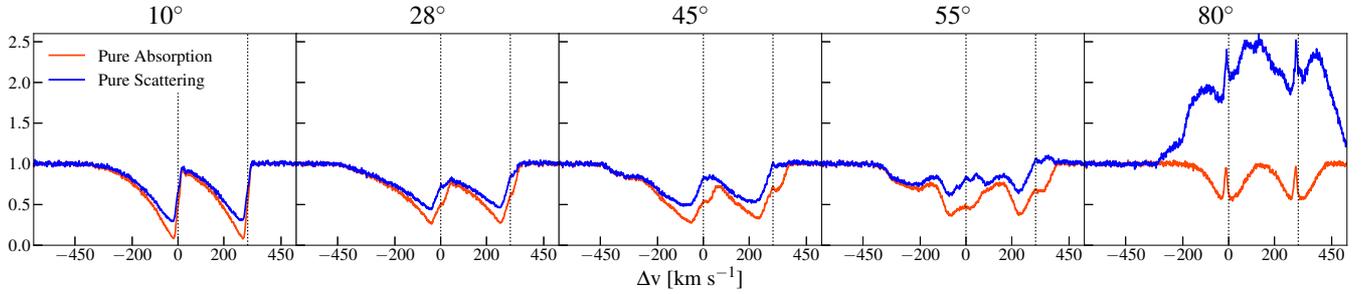}
\caption{Line profiles for the Model MHD1.  The pure scattering case results in strong emission at large inclinations. The pure absorption profiles are deeper and show only a slight emission feature at line center at high inclination.} 
\label{fig:mhdspect}
\end{figure*}

\begin{figure*}
\centering
\includegraphics[width=\textwidth]{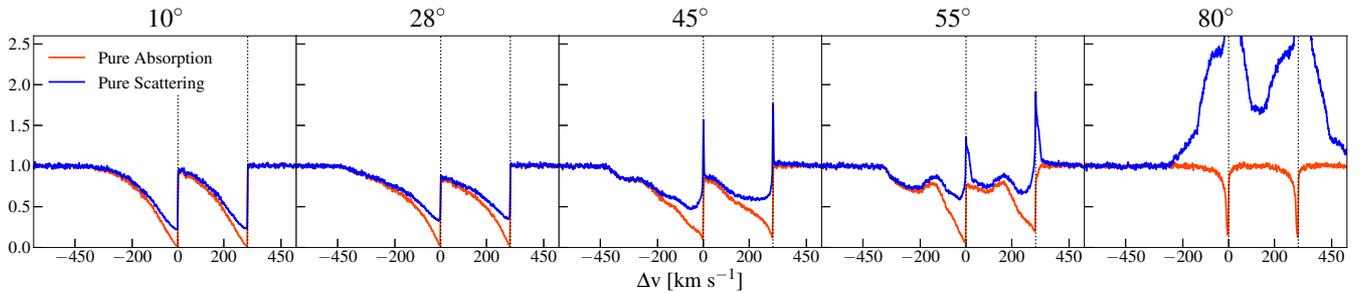}
\caption{Profiles for Model MHD2, which has the same parameters as Model 1 (Figure \ref{fig:mhdspect}), but with the rotational velocity set to zero.}
\label{fig:norot}
\end{figure*}

\begin{figure*}
\centering
\includegraphics[width=\textwidth]{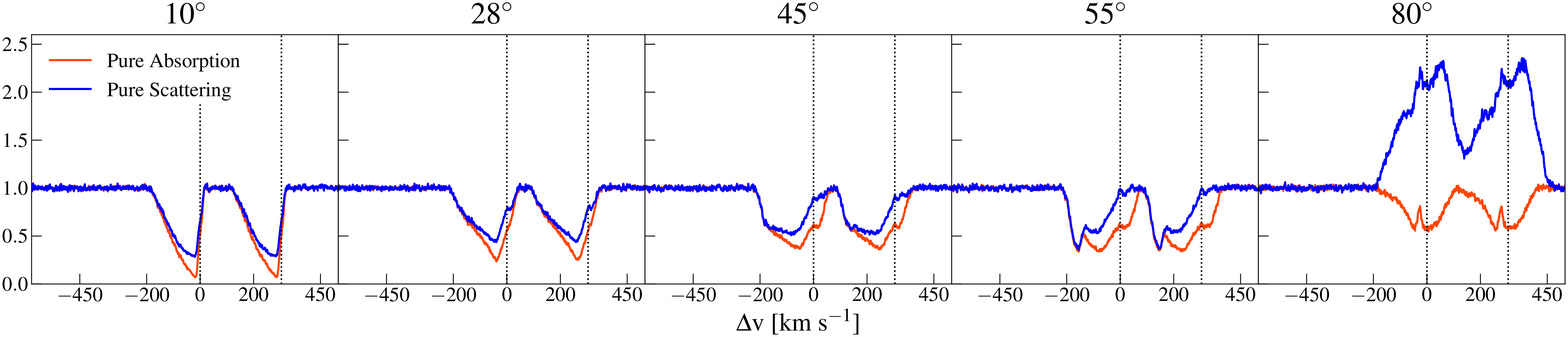}
\caption{Profiles for Genwind Model G3.}
\label{fig:genwind}
\end{figure*}

\section{Results}
\label{sec:results}

\subsection{Fiducial MHD model}

In Figures \ref{fig:mhdgrid1} and \ref{fig:mhdgrid2} we show the radial velocity, the azimuthal velocity, and the density of the MHD wind model.  The structure is such that the highest velocities are present in the wind regions closest to the inner cavity; the density contours are almost parallel to the
disc.

Figure \ref{fig:mhdspect} shows the model results for different inclination angles of the disc (and wind) axis to the line of sight.
The progression of the pure scattering profiles moving to higher inclinations is qualitatively similar to that found in previous disc wind simulations \citep[e.g.,][]{knigge95}; the photons that are scattered out of the line of sight at low inclinations are scattered into high-inclination lines of sight.  The pure absorption case obviously shows no such emission.

The low-inclination profiles are relatively smooth; more structure appears in the 
$i \geq 45^{\circ}$ profiles.  The roughly square
high velocity edge is due to the nature of the velocity law, which nearly constant above $\chi \gtrsim 10$, resulting in a pileup of absorbing material.  The total depth of this feature depends upon the extent of the grid; this is the result of making the unrealistic assumption of constant Na I abundance.  

The absorption profiles at $i = 45^{\circ}, 50^{\circ}$ also show a dip on the red side of the D2 line.  To investigate the origin of this feature, in Figure \ref{fig:norot} we show calculations for the same wind model but with $v_{\phi}$ set to
zero.  The comparison shows that this redshifted absorption is not due to the wind divergence but to rotation.  In general, the effect of rotation is to broaden the lines, most prominently at high inclinations as expected; the absorption at low inclinations is also weakened somewhat.

\subsection{Genwind models}

The model results shown in Figures \ref{fig:mhdspect} and \ref{fig:norot} are inconsistent with the strong, deep absorption over a wide range of velocities seen in V1057 Cyg and V1515 Cyg.  Assuming pure absorption only deepens the lines a modest amount. To explore whether varying the velocity field to a slower acceleration, we computed a Genwind model with the parameters shown in Figure \ref{fig:genwind}.  This model provides somewhat ``squarer'' absorption lines at low inclination, but again the pure scattering profiles are not sufficiently deep, especially at larger inclinations.

\begin{figure}
    \centering
    \includegraphics[width=0.4\textwidth]{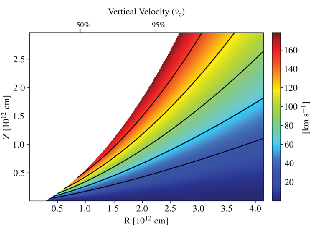}
    \caption{The $v_z$ velocity field of Genwind Model G3.}
    \label{fig:genwind3}
\end{figure}


\begin{figure*}
\centering
\includegraphics[width=\textwidth]{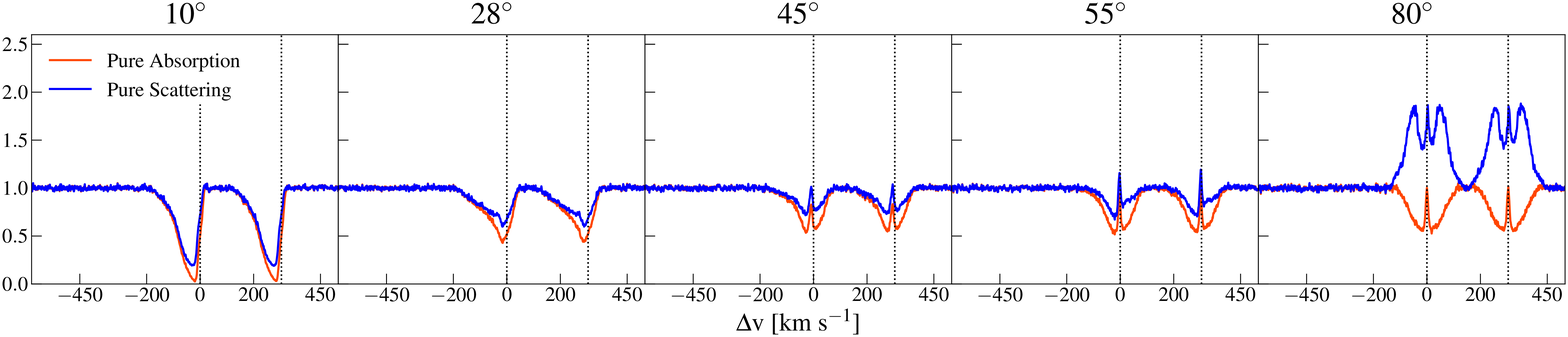}
\caption{Profiles for Genwind Model G4 (``vertical flow'').}
\label{fig:verticalspect}
\end{figure*}

The problems at low inclination are a result of the assumed wind geometry.
Figure \ref{fig:genwind3} shows the wind vertical velocity as a function of distance.  For reference we also mark the radial positions which encompass 50\% and 95\% of the total disc light at 5900 \AA.  This makes clear that, while low-inclination lines of sight pass through wind with low projected radial velocities, there is simply not enough absorbing material in the wind at intermediate-to-high velocities to ``cover'' the disc photospheric emission seen at low inclinations.

\subsection{"Vertical" Genwind model}
\label{sec:vertical}
The results of the previous model suggest that the cavity in the wind models might preclude having enough absorption seen at low inclinations.  To test this possibility,
we constructed a Genwind model with a very large value of $b$, which makes the flow nearly completely vertical from the disc. 
The resulting absorption profiles at $i =10^{\circ}$ are much deeper, in better agreement with the observations, demonstrating the importance of having absorbing material in what was previously a cavity (Figure \ref{fig:verticalspect}).  However, this model is not a very satisfactory solution, because with an essentially vertical wind there is very little blueshifted absorption at even modest inclinations.

\subsection{Genwind with inner wind models}

\begin{figure*}
    \includegraphics[width=0.4\textwidth]{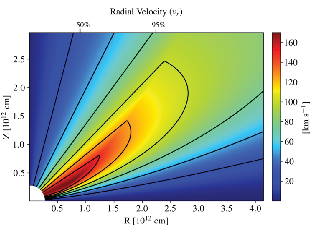}
    \includegraphics[width=0.4\textwidth]{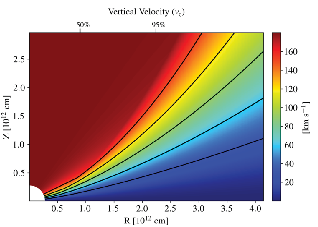}
    \caption{Radial and vertical velocity fields,
    for the Genwind Model G3 with an inner wind (Model G5).}
    \label{fig:stellarwindgrid}
\end{figure*}

\begin{figure*}
\centering
\includegraphics[width=\textwidth]{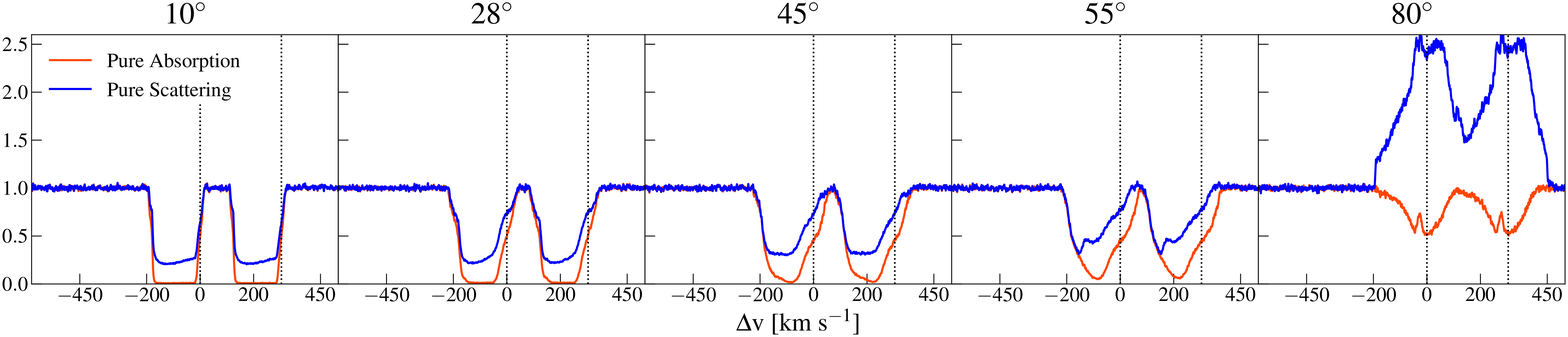}
\caption{Spectra for Genwind Model G3 + inner wind (Model G5)}
\label{fig:model5}
\end{figure*}

In addition to the unsatisfactory line profiles of the ``vertical''
wind at intermediate inclinations, 
it also seems physically unlikely that the wind near the disc would be essentially vertical, given the (at least initial) Keplerian rotation of the gas, which should tend to produce some outward flow due to centrifugal acceleration.  We then included
an inner wind to the Genwind Model G3,
as outlined in 
\ref{sec:innerwindmodels}.  Figure \ref{fig:stellarwindgrid} shows the wind velocity structure adding the inner wind to the model in Figure \ref{fig:genwind}.  The resulting line profiles, shown in Figure \ref{fig:model5},
are a marked improvement over the previous results, producing strong absorption over a wide velocity range at low inclinations.  The profiles at $i = 45^{\circ}$ and $55^{\circ}$
are also deeper than the corresponding results for other wind models, though there is still a problem with the pure scattering case.  In particular, the $i= 55^{\circ}$
is not as deep as seen in FU Ori (Figure \ref{fig:fuori}).  The comparison with the pure absorption case shows that this is not a problem due to insufficient geometrical ``covering'' of the disc photospheric emission. 


\subsection{MHD with inner wind}

\begin{figure*}
\centering
\includegraphics[width=\textwidth]{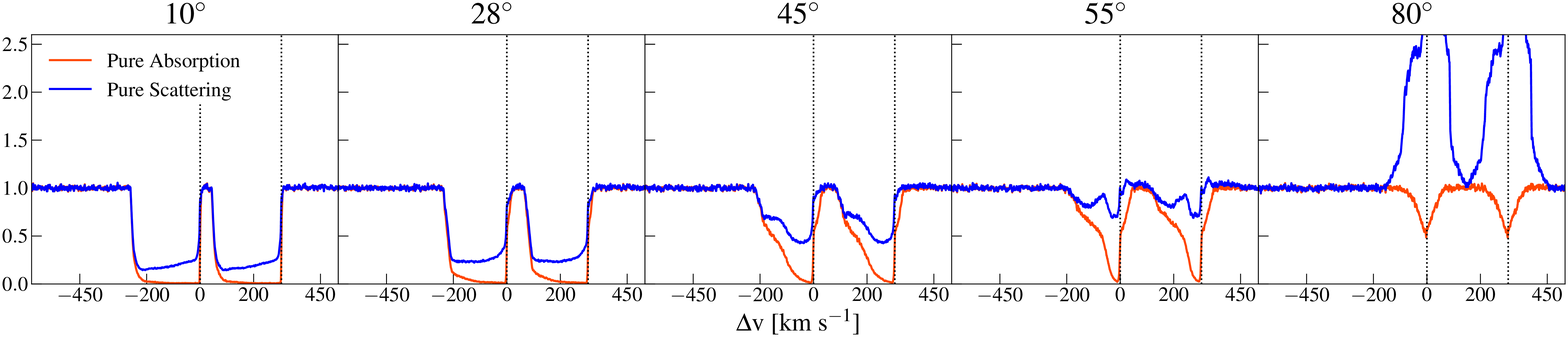}
\caption{Model MHD1i, which includes an inner wind.}
\label{fig:mhdinner}
\end{figure*}

Finally, we made a modified version of the inner wind solution to add to the fiducial MHD model.  We found that we needed to 
arbitrarily reduce the expansion velocities by a factor of two to make a reasonable comparison with the observations.  As shown in Figure \ref{fig:mhdinner}, 
the low-inclination profiles are reasonable.  If the source function could be modeled with pure absorption,
the profile at $i = 55^{\circ}$ would provide a good comparison with FU Ori, but once again the pure scattering profile is not deep enough.

\section{Discussion}

The streamlines of all but one of the wind geometries explored here exhibit significant divergence away from the rotational axis, as required for magnetocentrifugal acceleration of these cold flows to be important.  We have shown that such models, with an inner streamline at the stellar radius of a standard steady accretion disc,
cannot explain the Na I profiles for FU Ori objects seen nearly pole-on.  To explain the observations of such systems, we had to add an ``inner wind'' to fill in what would otherwise be a large central cavity in the flow.  We formulated this inner wind to have the same self-similar streamline shape as the outer wind for reasons of continuity and to avoid adding too many more adjustable parameters; however, real FU Ori winds
need not be self-similar.  Models exist in which vertical jets are driven by torsional Alfv\'en waves \citep[e.g.,][]{shibata85,hayashi96,goodson97}, resulting in highly-collimated flows such as our ``vertical wind'' Model G4. 
As shown in \S \ref{sec:vertical}, pure vertical winds cannot explain the large blueshifted absorption seen in non-pole-on objects, such as FU Ori; a hybrid, non-self-similar structure with a vertical inner wind combined with a divergent disc wind, such as the two-component outflows of \cite{goodson97}, could explain profiles for objects over a wide range of inclinations.  The structure of the inner wind is thus not well-constrained, especially because the temperature distribution in the inner disc regions are poorly understood (e.g., Figure \ref{fig:cartoon_fuori}).

The pure absorption models, though providing deeper line profiles in better agreement with observations,
are highly unlikely to be realistic.  For this limit to apply, that is to suppress scattering, collisional deexcitation would have to
exceed the radiative decay rate.  
The critical density for the D1 transition, assuming a temperature of 8000 K,
is $n_{ec} = A_{21}/q_{21} = 4 \times 10^{14} \cmthree$ \citep[using the collisional rate from][]{natta90}.
A typical wind density for the adopted velocities and mass loss rate here is $n_H \sim 10^{12} \cmthree$
(see wind grids) and the crude constraints of \cite{croswell87} require the wind to be sufficiently low in
temperature, with a modest level of hydrogen ionization $\sim 10^{-3}$, to avoid producing too 
much emission in H$\alpha$. Radiative trapping can help thermalize lines.  However, if we estimate the
escape probability as $P_{esc} \sim \tau^{-1}$, where $\tau$ is the line optical depth, then thermalization
would only occur for $\tau \sim A_{21}/(q_{21} n_e) \sim 4 \times 10^{14}/10^9 \gtrsim 10^5$, and for such
large optical depths one would expect to see damping wings in the Na I profiles.  

Thus, we expect the pure scattering case to be a good approximation for the Na I resonance lines.
The problem then becomes how to get deeper lines in better agreement with observations.  Observations of
the strong ultraviolet P Cygni line profiles in O stars frequently show deeper absorption than can
be explained with monotonic winds in the strict Sobolev approximation, which led \cite{lucy82,lucy83} to
suggest that non-monotonic velocity fields produce more backscattering and deepen absorption profiles
(see also \citealt{hamann81} for a discussion of the effects of turbulence).  As discussed in \S \ref{sec:obs},
there is extensive evidence for non-monotonic flows in FU Ori winds via absorption ``dips'', and theoretically
one might expect flows from the MRI-active disc to produce turbulent waves propagating outward.  We conjecture
that this non-monotonicity of the flows is why our smooth flows do not produce sufficiently deep absorption profiles.

Emission in very optically thick lines can be quenched by dust due to multiple scattering, although typical applications involve much stronger lines than the Na I resonance doublet \citep[e.g., Lyman $\alpha$;][]{ferland79}. While there is evidence for time-dependent dust formation in V1515 Cyg and in V1057 Cyg \citep{kenyon_kolo91, kolotilov97}, it seems likely that this dust formation occurs at large distances \citep{hartmann04}, particularly
as the inner wind, bathed by a $\sim$ 6000~K radiation field, is an inhospitable environment for forming enough dust to materially affect Na I line formation.

The observed Na I profiles of FU Ori and V1057 Cyg show
some slight redshifted absorption, extending to about
60-100 $\kms$ redward of line center (Figure \ref{fig:fuori}) (there is some evidence for a similar, though weaker and less redshifted, redshifted absorption component in V1515 Cyg).
As this is clearly seen in both members of the doublet
in FU Ori (and in other spectra of V1057 with less blueshifted wind absorption), these features are clearly
due to Na I.  Although some of the pure-absorption calculations also show similar features at intermediate inclination, the pure-scattering cases mostly show an
emission feature.  Our interpretation of the observed
redshifted dips is that they represent a combination of photospheric features and near-photospheric
absorption from the base of the rotating wind. Qualitatively
similar features for redshifted absorption were predicted for a simple rotating wind model by \cite{calvet93} (see bottom right two panels in their Figure 3).  The extent of the redshifted absorption in FU Ori is similar to the projected rotational velocity of photospheric lines,
supporting this interpretation
(see also discussion of Figures
\ref{fig:mhdspect} and \ref{fig:norot}). We note that 
the absorption in 
V1057 Cyg extends to larger redshifts than characteristic
of the photospheric lines, so simple rotation does not explain the observations.  However, higher rotational
velocities may be expected beyond local Keplerian values
if there is rotational magnetic acceleration (e.g., lower right panel of Figure \ref{fig:bpstd}).  (Note that the
\citealt{calvet93} models did not include any such spinup.)
In any case, our calculations of profiles at low velocities should not be taken as definitive, because we do not include a photospheric
profile, and because our treatment of low-velocity absorption is uncertain due to the questionable applicability of the Sobolev approximation.

One of the striking features of the pure scattering profiles is the appearance of emission features at high inclination.
This behavior is well-known from previous calculations of disc winds from cataclysmic variables \citep{shlosman1993,knigge95,long02,noebauer10,kusterer14,matthews15} and
AGN \citep{matthews16}.  The FU Ori objects
are generally surrounded by substantial dusty envelopes,
in some cases optically visible only due to viewing along
outflow holes \citep[e.g.,][]{kenyon91}, so observing
systems at inclinations as large as $80^{\circ}$ is unlikely, at least for Na I.  
However, there may well be objects that are visible at inclinations $\sim 60^{\circ}$ that could be studied to test the model
predictions.  In this regard, spectra of objects seen
at a variety of inclinations in scattered light along
outflow holes, such as seems to be the case for R Mon
\citep{jones82},
might reveal evidence for a progression from deep absorption to weaker absorption and even redshifted emission.

\section{Summary}

We have presented theoretical Na I resonance line profiles for parameterized disc winds for comparison with observations of the main (original) FU Ori objects.  We show that the deep, broad absorption P Cygni profiles cannot be explained easily with
models in which the disc wind diverges from the inner edge of the disc (as in Figure \ref{fig:diskwind_starx}), as this geometry cannot extinct enough of the optical disc photosphere at
intermediate-to high velocities at low inclination angles.  We also place weak constraints on the
divergence of the FU Ori wind. The assumption of pure scattering in the Na I resonance lines, which should be a reasonable approximation, makes it difficult to produce the extremely deep absorption seen in V1057 and V1515 Cyg;
turbulence and/or non-monotonic velocity fields may be
required, as might be expected for winds from turbulent
accretion discs.  The models also predict that redshifted emission components should become more prominent when the discs are viewed at higher inclination, which may be testable.  

We defer consideration of ionization balance to subsequent work which would require modeling other species to constrain wind temperatures
(\citealt{croswell87}; see also \citealt{natta90} for a discussion in the context of T Tauri winds).

The wind models presented here provide a highly preliminary view of what might be learned from modeling FU Ori winds.  As the inner discs of FU Ori systems are thought to be highly ionized, global ideal MHD simulations should eventually be able to make more detailed predictions for observed spectra.  However, our results suggesting that an inner disc wind is necessary means that inner boundary conditions for such simulations 
are likely to be critical to obtaining a satisfactory comparison with observed P Cygni profiles.  It also may be that the winds must
be turbulent to produce deep enough absorption in the Na I lines, such that the Sobolev approximation is not applicable for calculating the line profiles.

\section*{Acknowledgements}
We thank Lynne Hillenbrand for providing
the spectra shown in Figure 1.
This work was supported in part by the University of Michigan, Ann Arbor, including
the use of computational resources and services provided by its Advanced Research Computing unit.  The project was initiated at the Kavli Institute for Theoretical Physics, supported in part by the National Science Foundation under Grant No. NSF PHY17-48958. JHM is supported by the Science and Technology Facilities Council under grant ST/N000919/1. 
We would like to thank C. Knigge and S. A. Sim for helpful discussions. 
We also gratefully acknowledge the use of matplotlib \citep{matplotlib}.

\appendix

\section{Wind Divergence tests}

One of the original goals of this investigation was to try to constrain 
the divergence of the wind of FU Ori.  To that end we explored Genwind models with
different opening angles, as noted in Table 1 and as displayed in Figure \ref{fig:streamlines} in the Appendix.  The resulting profiles
are shown in Figures \ref{fig:model6}-\ref{fig:model7}. 
The pure absorption cases suggest that mode G7 has too
large a divergence, but scattering tends to fill in
the red-shifted part of the line profile.  Thus,
constraints on wind divergence will require further
understanding of the appropriate line source functions.

\begin{figure}
    \centering
    \includegraphics[width=0.4\textwidth]{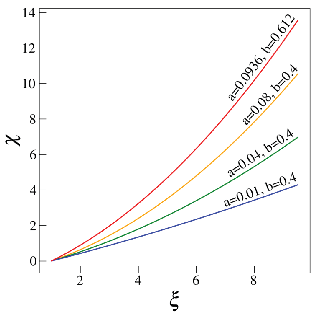}
    \caption{The streamlines for the model tests for the effect of divergence on the line profiles.  The upper curve is that of the MHD model.}
    \label{fig:streamlines}
\end{figure}

\begin{figure*}
\centering
\includegraphics[width=\textwidth]{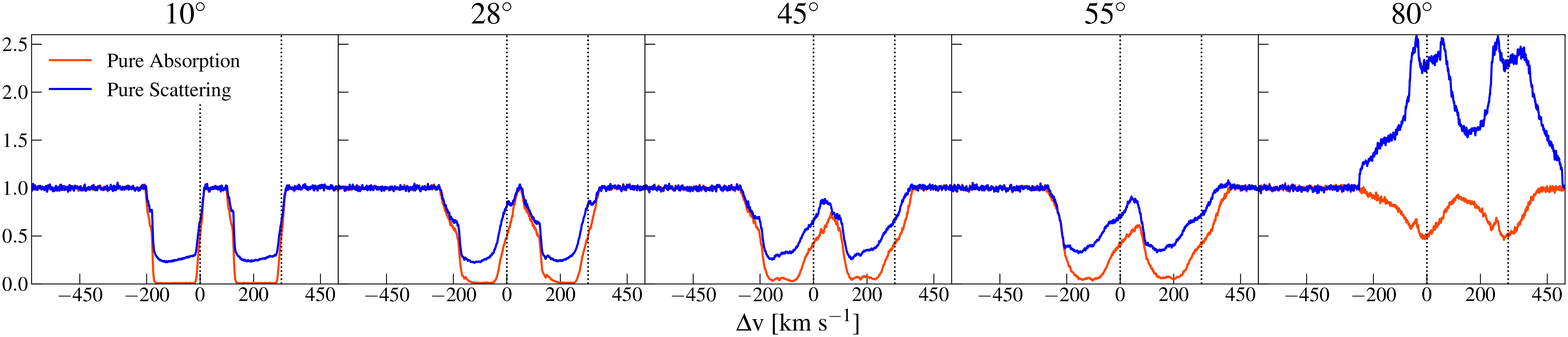}
\caption{Wind Model G6.}
\label{fig:model6}
\end{figure*}

\begin{figure*}
\centering
\includegraphics[width=\textwidth]{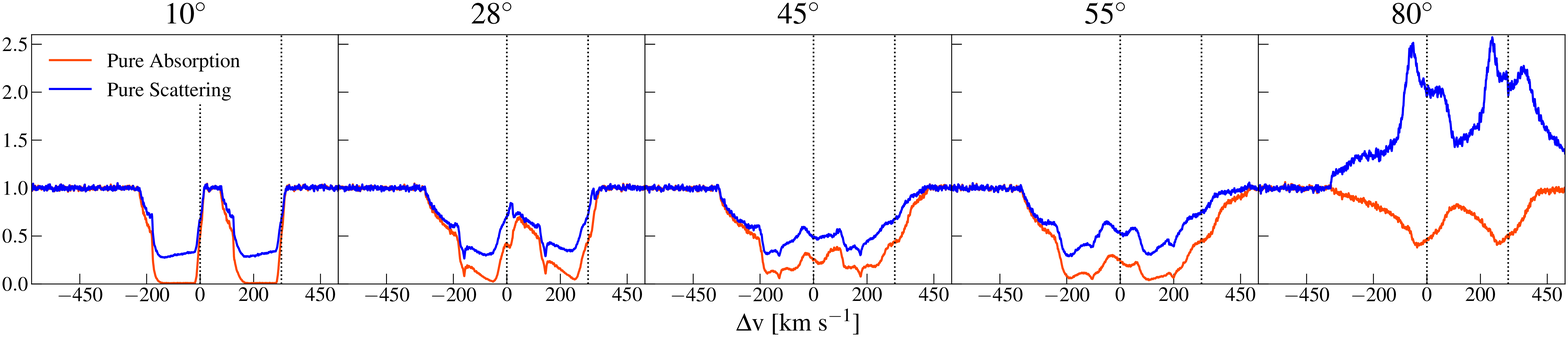}
\caption{Wind Model G7.}
\label{fig:model7}
\end{figure*}

%

\bibliographystyle{mnras}
\bibliography{refs}


\bsp    
\label{lastpage}
\end{document}